\documentclass[twocolumn]{svjour3}          % twocolumn
\smartqed  % flush right qed marks, e.g. at end of proof
\usepackage{graphicx}
\usepackage{hyperref}
\usepackage{amssymb}
\usepackage{xcolor}% http://ctan.org/pkg/xcolor
\newcommand{\mgamc}{{\tt MG5\_aMC}}

\newcommand{\parm}{\mathord{\color{black!33}\bullet}}%
\begin{document}

\title{On the maximal use of Monte Carlo samples:\\  re-weighting events at NLO accuracy}
%\thanks{IPPP/16/55, MCNET16.xx}
%\thanks{Grants or other notes
%about the article that should go on the front page should
%placed here. General acknowledgments should be placed at the end of the article.}

%\subtitle{Do you have a subtitle?\\ If so, write it here}

\author{Mattelaer Olivier
%        \and
%        Second Author\thanksref{e2,addr2,addr3} %etc.
}

%\thankstext[$\star$]{t1}{Thanks to the title}
%\thankstext{e1}{e-mail: o.p.c.mattelaer@durham.ac.uk}
%\thankstext{e2}{e-mail: magic2@xxx.xx}

\institute{Institute for Particle Physics Phenomenology (IPPP) \at
 Durham University,\\
 South Road 1Durham,\\ DH1 3LF, England.\\
  \email{o.p.c.mattelaer@durham.ac.uk}
%          \and
%         Second Address, Street, City, Country\label{addr2}
%         \and
%          \emph{Present Address:} Street, City, Country\label{addr3}
}

\date{Preprint: IPPP/16/55, MCNET-16-26}
% The correct dates will be entered by the editor

\maketitle

\begin{abstract}
Accurate Monte Carlo simulations for high-energy events at CERN's Large Hadron Collider, are very expensive, both from the computing and storage points of view. We describe a method that allows to consistently re-use parton-level samples accurate up to  NLO in QCD  under different theoretical hypotheses. We implement it in MadGraph5\_aMC@NLO and show its validation by applying  it to several cases of practical interest for the search of new physics at the LHC. 

\keywords{MadGraph5\_aMC@NLO \and NLO \and re-weighting \and automation}
% \PACS{PACS code1 \and PACS code2 \and more}
% \subclass{MSC code1 \and MSC code2 \and more}
\end{abstract}

\section{Introduction}

The search of new physics is one of the main priorities of the LHC. The recent observation of an anomaly in the di-photon spectra \cite{aa_atlas,CMS:2015dxe} gives hope that we might have a first evidence of Beyond Standard Model (BSM) physics very soon. In that case, we would only be at the beginning of a long program of investigations of what the underlying physics is. In any case, searches of new particles or modifications of the interactions among the SM particles will continue as well as progress associated to our ability to provide precise predictions to be compared with data.

In the recent years,  efforts have focussed on providing accurate theoretical predictions for a large number of BSM models at Leading Order (LO), in the form of event generators. First, various programs such as FeynRules \cite{Alloul:2013bka}, LanHep \cite{Semenov:2014rea} or Sarah \cite{Staub:2013tta} have automated the extraction of the Feynman rules from a given Lagrangian.
 Secondly several matrix element based generators like MadGraph5\_\-aMC\-@NLO \cite{Alwall:2014hca} (referred to as \mgamc\ later on), Sherpa \cite{Gleisberg:2008ta} or Whizard \cite{Kilian:2007gr} have extended the class of BSM model they support with extensions in various directions: high spins, high color representations and any kind of Lorentz structure \cite{Degrande:2011ua,deAquino:2011ub,Hoche:2014kca}. More recently, automated Next-to-Leading Order (NLO) prediction (in QCD) for BSM models are available thanks to the  NLOCT \cite{Degrande:2014vpa} package of FeynRules which adds in the model the additional elements ($R2$ and UV counter-terms) required by loop computations. 

It is now possible to generate Monte-Carlo sample for a large class of BSM theories at LO and for an increasing number at NLO accuracy. Even though technically possible, producing samples for many models and benchmark points down to full detector level at the high luminosity expected at the LHC would require an unmanageable number of computing and storage resources. However, the stages of a simulation (parton-level generation, parton-shower and hadronisation, detector simulation, and reconstruction) are independent and factorise. Therefore changes in local probabilities happening at very short distance, i.e. from  BSM physics,  decouple from the rest of the simulation stages. This is particularly interesting since the slowest part of the simulation is the full simulation of the detector.

A logical possibility therefore arises: one can generate large samples under a SM or basic BSM hypothesis and then continuously and then locally deform the probability functions associated to the distributions of parton-level events in the phase space by changing the ``weight" of each event in a sample to account for an alternative theory or benchmark point. Under a not-too-restrictive set of hypotheses which are easy to list, such an event-by-event re-weighting can be shown to be exactly equivalent (at least in the infinite statistic limit) to a direct generation in the BSM. Note that such an event-by-event re-weighting is conceptually different from the very common yet very crude method where events are re-weighted using a pivotal one-dimensional distribution. Event-by-event re-weighting is a common practice in MC simulations, yet currently it has been only publicly available at LO \cite{Gainer:2014bta,Baglio:2014uba} or available at NLO for very specific cases (e.g.  \cite{Frixione:2002ik}) or in methods where NLO accuracy is far from ensured \cite{Alioli:2010xd,REPOLO}. 
 It is the aim of this work to show that a consistent  (and practical) re-weighting of events can also be done at NLO accuracy.

The plan of this paper is as follows. Before introducing the NLO re-weighting method, we will focus on the LO case in order to explain the intrinsic limitations of such types of methods (Section \ref{sec:LO}).
In Section \ref{sec:NLO}, we present three types of NLO re-weighting, two of them correspond to methods already introduced in the literature  \cite{Baglio:2014uba,Frederix:2014hta}. The third one is the NLO accurate re-weighting method introduced here for the first time.  In Section \ref{sec:validation}, we present some validation plots performed with \mgamc. We then present our conclusions in Section \ref{sec:conclusion}.

\section{Re-weighting at the leading order}
\label{sec:LO}

As stated in the introduction, the re-weighting method consists in attaching a new weight to every parton-level event as corresponding to a different scenario. The new weights allow to predict accurately (up to statistical precision) all the LO differential distributions at the parton level, leading also to the possibility of performing a single shower and detector simulation for all the models under consideration. At LO accuracy the new weight ($W_{new}$) can be easily obtained from the original one ($W_{orig}$) by simply multiplying it by the ratio of the matrix-elements estimated on that event for both models (noted respectively $|M_{orig}|^2$ and $|M_{new}|^2$) \cite{Gainer:2014bta,Baglio:2014uba}:
\begin{equation}\label{me_re-weighting}
W_{new} = \frac{|M_{new}|^2}{|M_{orig}|^2}W_{orig}.
\end{equation}
In practice, in a weighted Monte-Carlo generation, the weights are simply given by\footnote{\label{note1}For the simplicity of the discussion, we will always consider that the sum of the weights is equal to the total cross-section of the sample. }
\begin{equation}\label{me_re-weightingl}
W_{orig} =  f_1(x_1,\mu_F)\cdot f_2(x_2,\mu_F)\cdot|M_{orig}|^2\cdot \Omega_{PS}\,,
\end{equation}
where $f_i(x_i,\mu_F)$ is the parton-distribution function estimated on the Bjorken fraction $x_i$ at the factorization scale $\mu_F$. $\Omega_{PS}$ is the phase-space measure of the phase-space volume associated to the events.\footnote{The normalisation choice implies that the phase-space factor $ \Omega_{PS}$ is proportional to $N^{-1}$ where $N$ is the number of phase-space points used to probe the phase space. }
% This factor depends both of the phase-space parametrization and of the total number of phase-space points used to build the estimator of the cross-section.
From this equation it is clear that Eq. \ref{me_re-weighting} is the correct procedure since the weight is exactly multiplicative. This property is preserved by the unweighting procedure making Eq. \ref{me_re-weighting} to hold for both weighted and un-weighted samples
 (an actual proof is presented in Appendix \ref{annexe:proof}).

A few remarks are in order regarding the range of validity of this method. First, even if the method returns the correct weight, it requires that the event sampling related to $W_{orig}$ covers appropriately the phase-space for the new theory. In particular, $W_{orig}$ must be non-zero in all regions where $W_{new}$ is non-vanishing. Though obvious, this requirement is in fact the most important and critical one. 
In other words, the phase-space where the new theoretical hypothesis contributes should be a subset of the original one. For example, re-weighing can not be used for scanning over different mass values of the final state particles\footnote{For intermediate particle a small variation of the mass --order of the width-- is reasonable.}, yet it is typically well-suited for probing different types of spin and/or coupling structures.
More in detail,  when the new theory has large contribution in a region of the phase-space where 
the original sample has only few events --since the original is sub-dominant in that part of the phase-space--, 
 the statistical uncertainty of the re-weighted sample becomes very large and the resulting predictions unreliable.
To appreciate quantitatively such an effect, we can use a naive estimator assuming a gaussian behavior. In that case one can write the estimated uncertainty as
\begin{equation} \label{eq:stat_uncertainty}
\Delta \mathcal{O}_{new} = \bar w\cdot \Delta  \mathcal{O}_{orig} + Std(w) \cdot \mathcal{O}_{orig}\,,
\end{equation}
where $\bar w$ and $Std(w)$ are respectively the mean and the standard deviation of the ratio of the weights and $ \mathcal{O}_{\parm}$, $\Delta  \mathcal{O}_{\parm}$ are an observable and the associated statistical uncertainty. As a consequence, the relative uncertainty can be enhanced if the weights have a large variance.
In Appendix \ref{annexe:proof}, we introduce, as a proof of principle, a second method on how to estimate the statistical uncertainty from the distribution of the weights.

Second, the parton-level configuration feeder to parton-shower programs not only depends of the four-momenta but also of additional information, which is commonly encoded in the LesHouches Event File (LHEF) \cite{Alwall:2006yp,Andersen:2014efa}. Consequently, re-weighting  by an hypothesis that does not preserve such  additional information is not accurate. In general, such informations are related to:

\begin{itemize}
\item {\bf Helicity:} The helicity state of the external states of a parton-level event is optional in the LHEF convention, yet some programs (e.g. \cite{Meade:2007js}) use this information to decay the heavy state with an approximated spin-correlation matrix.  In this case it is easy to  modify Eq.~\ref{me_re-weighting} to correctly take into account the helicity information by using the following re-weighting:
\begin{equation}\label{me_re-weighting_hel}
W_{new} = \frac{|M^{h}_{new}|^2}{|M^{h}_{orig}|^2}W_{orig}\,,
\end{equation}
where $|M^{h}_{new}|^2$ and $|M^{h}_{orig}|^2$ are the matrix elements associated to the event for a given helicity $h$ --the one written in the LHEF-- and for the corresponding theoretical hypothesis. This re-weighting is allowed since the total cross-section is equal to the sum of the individual polarized cross-sections. 

\item {\bf Color-flow:} A second piece of information presented in the LHEF is the color assignment in the large $N_c$ limit. This information is used as the starting point for the dipole emission of the parton shower and therefore determines the result of the QCD evolution  and hadronisation.  Such information is untouched by the re-weigthing limiting the validity of the method.
For example, it is not possible to re-weight events with a Higgs boson, with a process where the Higgs boson is replaced by a colored particle. One could think that, as for the helicity case, one could amend the re-weighting formula to be able to handle modifications in the relative importance between various flows. While possible in principle, in practice 
such re-weighting would require to store additional  information (the relative probabilities of all color flows in the old model) in the LHEF, something that does not seem practical. 

\item {\bf Internal resonances:} In presence of on-shell propagators, the associated internal particle is written in the LHEF. This is used by the parton-shower program to guarantee that the associated invariant mass is preserved during the re-shuffling procedure intrinsic to the showering process. Consequently, modifying the mass/width of internal propagator should be done with caution since it can impact the parton-shower behaviour. This information can not be corrected via a re-weighting formula, as it links in a non-trivial way short-distance with long-distance physics.
\end{itemize}

Selected results obtained with this re-weighting are presented in Section \ref{sec:validation}.

\section{Next to leading order re-weighting}
\label{sec:NLO}

In this section, we will present three re-weighting methods for NLO samples. First we will present a LO type of re-weighting that we dubbed ``Naive LO-like'' re-weighting  introduced in {\tt VBFNLO} (i.e. {\tt REPOLO} \cite{REPOLO})  and {\tt MadSpin} \cite{Artoisenet:2012st,Frixione:2007zp}. As it will become clear later, this method is not NLO accurate and should be used only if the difference between the two theories factorizes from the QCD production. The second method that we propose is original and consists in a fully accurate and general NLO re-weighting. Finally, we present the ``loop-improved'' re-weighting method \cite{Frederix:2014hta} to perform approximate NLO computation for loop-induced processes when the associated two-loop computations are not available.

\subsection{Naive LO-like re-weighting}

Following the {\tt MC@NLO} method \cite{Frixione:2010wd}, the cross-section can be decomposed in two parts, each of which can be used to generate events associated to a given final state multiplicity:
\begin{eqnarray}
d\sigma^{(\mathbb{H})} &=& d\sigma^R - d\sigma^{MC},\nonumber\\
d\sigma^{(\mathbb{S})} &=& d\sigma^{MC} + \sum_{\alpha=S,C,SC} d\sigma^\alpha,\label{eq:s/h_event}
\end{eqnarray}
where $R, S, C, SC, MC$
correspond respectively to the contributions of the fully-resolved configuration (the real),  of its soft, collinear, soft-collinear limits (the counter-events) and the Monte-Carlo (MC) counter-term. The $(\mathbb{S})$ (for standard) part corresponds to events generated with the Born configuration (N particles in the final state), while the  $(\mathbb{H})$ (for hard) part corresponds to events generated with the real configuration (N+1 particles in the final state). The MC counter-term (shower dependent) assures the coherent treatment with the parton-shower (no double counting) while preserving the NLO accuracy of the computation.

The Naive LO-like re-weighting computes the weights  based on the multiplicity of the events before parton shower. i.e.,
\begin{eqnarray}
    W^{(\mathbb{S})}_{new} &=& \frac{\mathcal{B}^{new}}{\mathcal{B}^{orig}}  W^{\mathbb{S}}_{orig},\\
    W^{(\mathbb{H})}_{new} &=& \frac{\mathcal{R}^{new}}{\mathcal{R}^{orig}}  W^{\mathbb{H}}_{orig}.     
\end{eqnarray}
$W^{(\mathbb{S})}_{\parm}$, $W^{(\mathbb{H})}_{\parm}$ are respectively the weights for Born/real topology events.  $\mathcal{B}^{\parm}$ is the Born matrix element squared ($ |M_n^{\parm}|^2$) while
 $\mathcal{R}^{\parm}$ is the real matrix element squared  ($|M_{n+1}^{\parm}|^2$).

As this method does not consider the dependence of the virtual contributions, it fails to be NLO accurate.  To ensure NLO accuracy, it requires that the effect of the new theory factorises out,  i.e.,  when
 \begin{equation}
  \frac{\mathcal{B}^{new}}{\mathcal{B}^{orig}} = \frac{\mathcal{V}^{new}}{\mathcal{V}^{orig}} = \frac{\mathcal{R}^{new}}{\mathcal{R} ^{orig}}= \textrm{Cst}
\end{equation}
 where $\mathcal{V}^{\parm}$ is the finite piece of the virtual contribution (the interference term between the Born and the loop amplitude). Such relation should hold over the full phase-space with a universal constant since the MC counter terms connect the born 
and the real in a non local way.  Nevertheless, as we will see later, the effect of the MC counter terms are quite mild, as expected since their contribution to the total cross-section are exactly zero by construction. This allows the Naive LO-like method to nicely approximate the NLO differential cross-section for many processes/theories where the last equation needs to be valid only phase-space point by phase-space point (i.e. when the ratio of the real matches the ratio of the Born and of the virtual in the soft and/or collinear limit).
 
\subsection{NLO re-weighting}

In order to have an accurate NLO re-weighting method, one should explicitly factorise out the dependence in the (various) matrix elements (i.e. in the Born squared matrix element --$\mathcal{B}$-- ,
the real squared matrix element --$\mathcal{R}$-- and in the finite piece of the virtual --$\mathcal{V}$--).
 We use the decomposition of the differential described in \cite{Frederix:2011ss}\footnote{We also use the same (MC) counter terms as described in that paper.} introduced in the context of  the evaluation of the systematics uncertainties:
\begin{eqnarray}
 d\sigma^\alpha&=&f_1(x_1,\mu_F)f_2(x_2,\mu_F) \left[\mathcal{W}^\alpha_0 +\mathcal{W}^\alpha_F \,\textrm{log}\left(\mu_F/Q\right)^2 + \right.\nonumber\\
    &&\left.\mathcal{W}^\alpha_R \,\textrm{log}\left(\mu_R/Q\right)^2 \right] d\chi^\alpha, \label{eq:decomp}
\end{eqnarray}
where the $\alpha$ index is either $R, S, C, SC, MC$ (see previous sub-section).
% $f_1(x_1, \mu_F)$, $f_2(x_2, \mu_F)$ are the parton distribution evaluated on the associated bjorken fraction $x_1$ and $x_2$ at the factorization scale $\mu_F$,
$Q$ is the Ellis-Sexton scale and $d\chi^\alpha$ is the phase-space measure.

The expression of the $\mathcal{W}^\alpha_0$, $\mathcal{W}^\alpha_F$, $ \mathcal{W}^\alpha_R$ are given in the appendix of \cite{Frederix:2011ss} and are not repeated here.
All those expressions have linear dependencies in the Born, the virtual, the real and the color connected Born $\mathcal{B}_{CC}$ (this term is defined in Eq. (3.24) of \cite{Frederix:2009yq}).
This allows us to decompose the corresponding  expressions as:\footnote{Due to the presence of multiple couter terms, the kinematic configuration on which the matrix-element is evaluated is not unique: 
an implicit sum over such kinematical configurations is assumed here and in the rest of the paper.}
\begin{eqnarray}
\mathcal{W^\alpha_\beta} &=\phantom{+}& \mathcal{B}*\mathcal{C}^\alpha_{\beta,B} +\mathcal{B}_{CC}*\mathcal{C}^\alpha_{\beta,B_{CC}} \nonumber \\
&\phantom{=}+& \mathcal{V}*\mathcal{C}^\alpha_{\beta,V} + \mathcal{R}*\mathcal{C}^\alpha_{\beta,R}
\end{eqnarray}
where the $\beta$ index is either  $0$, $R$ or $F$. The $\mathcal{C}^\alpha_{\beta,\parm}$ are expressions which do not depend of either the PDF/scale or the matrix-element.
From this expression we define the following three terms:\footnote{One can notice that $\mathcal{W}_{\beta,V}^{\alpha} = \mathcal{W}_{\beta,R}^{\alpha} = 0 $ for $\beta = R,F$ due to the use of the Ellis-Sexton scale \cite{Alwall:2014hca}.  }
\begin{eqnarray}
\mathcal{W}^\alpha_{\beta,B} &\equiv& \mathcal{B}*\mathcal{C}^\alpha_{\beta,B} +\mathcal{B}_{CC}*\mathcal{C}^\alpha_{\beta,B_{CC}}, \\
\mathcal{W}^\alpha_{\beta,V} &\equiv&\mathcal{V}*\mathcal{C}^\alpha_{\beta,V},\\
\mathcal{W}^\alpha_{\beta,R} &\equiv& \mathcal{R}*\mathcal{C}^\alpha_{\beta,R}.
\end{eqnarray}

%& \equiv\phantom{+}& \mathcal{W}^\alpha_{\beta,B} + \mathcal{W}^\alpha_{\beta,V} + \mathcal{W}^\alpha_{\beta,R}\,,
% Besides, it defines the $\mathcal{W}^\alpha_{\beta,\parm}$  object which have a linear dependence in either $\mathcal{B}$, $\mathcal{V}$ or $\mathcal{R}$.

By keeping track of the  $\mathcal{W}^\alpha_{\beta,\parm}$ at the generation time and writing it in the final event, one can perform an NLO re-weighting by: 
\begin{eqnarray}
  \mathcal{W}_{\beta,B}^{\alpha,new} &=& \frac{\mathcal{B}^{new}}{\mathcal{B}^{old}} * \mathcal{W}_{\beta,B}^{\alpha,old}, \nonumber\\
  \mathcal{W}_{\beta,V}^{\alpha,new} &=&\frac{\mathcal{V}^{new}}{\mathcal{V}^{old}} * \mathcal{W}_{\beta,V}^{\alpha,old}, \nonumber \\
  \mathcal{W}_{\beta,R}^{\alpha,new} &=& \frac{\mathcal{R}^{new}}{\mathcal{R}^{old}} * \mathcal{W}_{\beta,R}^{\alpha,old}. \label{eq:simplenlorwgt}
\end{eqnarray}
The final weight associated to the event can then be calculated by combining those various pieces as it is done  for the estimation of the systematics uncertainty (see Appendix of \cite{Frederix:2011ss}).
One can notice that the color-connected Born is simply re-weighted by the ratio of the Born which can lead to a breaking of the NLO accuracy of the method. However such an approximation does not consist in an additional limitation of the method since the re-weighting factors should differ only if the two theories present a difference in the relative importance of the various color-flows (a case already not handled at LO accuracy).

More generally, the possible drawbacks and limitations on the statistical precision of the method are the same as for the LO case. However, for NLO calculations in \mgamc\ we face one additional source of statistical uncertainty due to the method used to integrate the virtual contribution. This method reduces the number of computations of the virtual by using an approximate of the virtual contribution based on the Born amplitudes times a fitted parameter $\kappa$. It performs a separate phase-space integration to get the difference between the virtual and its approximation (full description of the method is presented in Section 2.4.3 of \cite{Alwall:2014hca}). Schematically it can be written as:
\begin{equation} 
\int (\mathcal{B}+\mathcal{V}) = \int (\mathcal{B}+\kappa \mathcal{B}) \,\,+\,\, \int (\mathcal{V}-\kappa \mathcal{B}). \label{eq:virttrick}
\end{equation}
If it exists a value of $\kappa$ such that $\kappa \mathcal{B}\approx \mathcal{V}$, the second integral is approximately zero and does not need to be probed as often as the first integral (thanks to importance sampling \cite{Weinzierl:2000wd}), reducing the amount of time used in the evaluation of the loop-diagrams. However the re-weighting proposed in Eq. \ref{eq:simplenlorwgt} will highly enhance the contribution of the second integral since each term of the integral will be re-weighted by a different factor, having a direct impact on the statistical uncertainty.

To reduce this effect, we propose to use a slightly more advanced re-weighting technique. We split the contribution proportional to the Born ($\mathcal{W}^\alpha_{\beta,B}$) in two parts: $\mathcal{W}^\alpha_{\beta,BC}$ and $\mathcal{W}^\alpha_{\beta,BB}$. $\mathcal{W}^\alpha_{\beta,BC}$ is the part, proportional to the Born, related to the one of the counterterms, while $\mathcal{W}^\alpha_{\beta,BB}$ includes all of the other contributions (the Born itself and the approximate virtual). We then apply the following re-weighting:
\begin{eqnarray}
   \mathcal{W}_{\beta,BB}^{\alpha,new} &=& \frac{(\mathcal{B}^{new}+\mathcal{V}^{new})}{(\mathcal{B}^{old}+\mathcal{V}^{old})} * \mathcal{W}_{\beta,BB}^{\alpha,old} \nonumber\\
   \mathcal{W}_{\beta,BC}^{\alpha,new} &=& \frac{\mathcal{B}^{new}}{\mathcal{B}^{old}} * \mathcal{W}_{\beta,BC}^{\alpha,old}  \nonumber\\
   \mathcal{W}_{\beta,V}^{\alpha,new} &=& \frac{(\mathcal{B}^{new}+\mathcal{V}^{new})}{(\mathcal{B}^{old}+\mathcal{V}^{old})} * \mathcal{W}_{\beta,V}^{\alpha,old}  \nonumber\\
      \mathcal{W}_{\beta,R}^{\alpha,new} &=& \frac{\mathcal{R}^{new}}{\mathcal{R}^{old}} * \mathcal{W}_{\beta,R}\,,^{\alpha,old} \label{eq:NLO_VIRT}
\end{eqnarray}
Both the virtual and the approximate virtual are re-weighted by the same pre-factor which should allow to limit the enhancement of the second integral.
The demonstration that such re-weighting is NLO accurate is presented in appendix~\ref{annexe:proof}. It can be intuitively understood considering ($\mathcal{B}+\mathcal{V}$) as a single block which is re-weighted accordingly. 

%Another difference with the LO production is the fact that NLO computation features negative events. While not changing the procedure in any way, it means that the unweighting is done such that 
%the sum of the absolute weight is preserved. In practise this means that the sum of the weight of the events is not equal to the actual cross-section (it is up to statistical fluctuation). 
%This means that the trivial re-weighting (both hyppothesis identical) will not return exactly the same cross-section. 

\subsection{Loop improved re-weighting}

A third type of re-weighting was originally introduced in the context of multiple Higgs production \cite{Frederix:2014hta,Maltoni:2014eza,Frederix:2016cnl}, which we now briefly describe. In this case the idea is to perform the NLO computation in the infinite top-mass limit and then re-introduce the finite top-mass effects via re-weighting. Eq. \ref{eq:NLO_VIRT} is directly applicable if the exact finite virtual part is known. If not,  one can still use  an approximate method:
\begin{eqnarray}
  \mathcal{W}_{\beta,B}^{\alpha,new} &=& \frac{\mathcal{B}^{new}}{\mathcal{B}^{old}} * \mathcal{W}_{\beta,B}^{\alpha,old}, \nonumber\\
  \mathcal{W}_{\beta,V}^{\alpha,new} &=&\frac{\mathcal{B}^{new}}{\mathcal{B}^{old}} * \mathcal{W}_{\beta,V}^{\alpha,old}, \nonumber \\
  \mathcal{W}_{\beta,R}^{\alpha,new} &=& \frac{\mathcal{R}^{new}}{\mathcal{R}^{old}} * \mathcal{W}_{\beta,R}^{\alpha,old}. \label{eq:loop-improved}
\end{eqnarray}
Both this method and the Naive LO-like method are not NLO accurate. However one can expect that the loop improved method has a better accuracy than the other one due to the correct treatment of the various counter terms.

\section{Implementation and validation}
\label{sec:validation}
The various methods of re-weighting discussed in the previous section have been implemented in \mgamc\  and are publicly available starting from version  2.4.0.
At the LO, the default re-weighting mode is based on the helicity information present in the event (Eq. \ref{me_re-weighting_hel}), while for NLO samples,
the default re-weighting mode is the NLO accurate one (Eq.~\ref{eq:NLO_VIRT}). Fixed-order NLO generation can not be re-weighted since no event generation is performed in this mode. 
A manual of the code is available online at the following address:\\ \href{http://cp3.irmp.ucl.ac.be/projects/madgraph/wiki/Reweight}{cp3.irmp.ucl.ac.be/projects/madgraph/wiki/Reweight}.

 In this section, we will present four validation examples covering the various types of re-weighting introduced in the previous section.  Since the purpose of this section is mainly to validate our method, the details of the simulation used (cuts, type of scale, ...) are kept to a minimum. Otherwise stated,  the settings used correspond to the default value of \mgamc\ (version 2.4.0).
 
\subsection{$ZW$ associated production in the Effective Field Theory at the LO}

For the first validation, we will use the Effective Field Theory (EFT) in the Electro-Weak sector \cite{Degrande:2012wf}. We will focus on the associated production of the $W$ and $Z$ boson for the following dimension six operator:
\begin{equation}
\mathcal{O}_{3W} = Tr\left[W_{\mu\nu}W^{\nu\rho}W_{\rho}{}^{\mu}\right],
\end{equation}
with
\begin{equation}
W_{\mu\nu}  = \frac{i}{2} g_W \tau^I (\partial_\mu W^I_\nu - \partial_\nu W^I_\mu
        + g_W \epsilon_{IJK} W^J_\mu W^K_\nu )
\end{equation}
and $g_W$ is the weak gauge coupling, $\tau^I$ are the pauli matrices and $W^I_\mu$ is the gauge Field of $SU(2)$.

\begin{figure}[ht]
\begin{center}
\begin{minipage}{1.0\linewidth}
\centering
\includegraphics[width=1.0\linewidth]{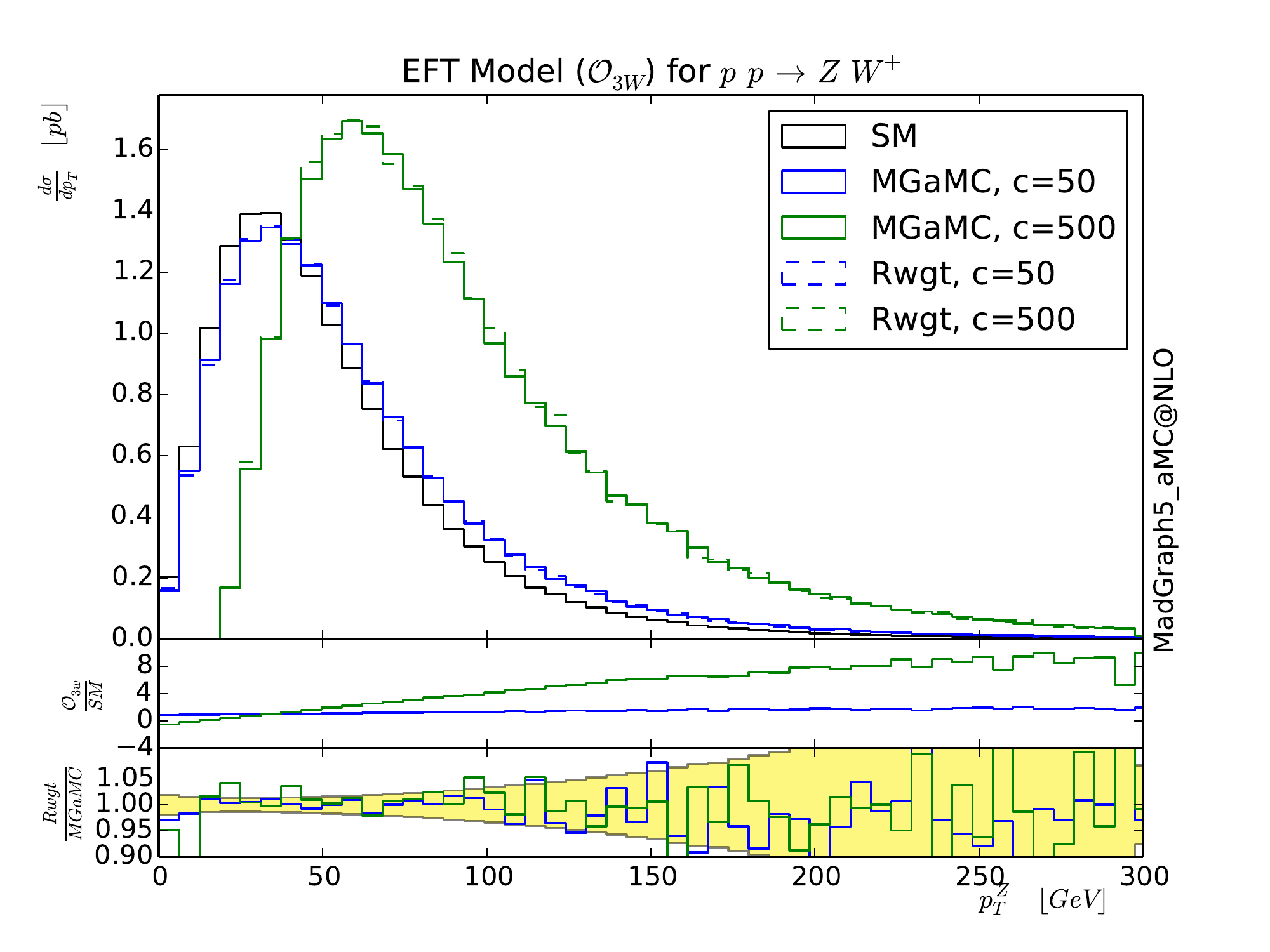}
\caption{\small{Differential cross-section for $p p \to Z W^+$ at 13 TeV LHC. This correspond to the Standard model plus the operator $\mathcal{O}_3W$  for two different couplings value. Only the SM contribution plus the interference term is kept on this plot. See text for details.
}}
\label{dim6_re-weighting_figure}
\end{minipage}\hfill
\end{center}
\end{figure}

In  Figure~\ref{dim6_re-weighting_figure} we present the differential distributions for the transverse momenta of the $Z$ boson at LO accuracy.
Starting from a sample of  Standard Model events (black solid curve), we have re-weighted our sample to get the SM plus the interference term with the dimension six operator for two values of the associated coupling:
 $c=50\, \textrm{TeV}^{-2}$ (dashed blue) and $c=500\, \textrm{TeV}^{-2}$ (dashed green). This second value is clearly outside the validity region for the EFT approach as the differential distributions turns to be negative at low transverse momentum. Nevertheless, having such large effects is interesting for the validation of the re-weighting method. The same differential distributions are generated with \mgamc\ (solid green and blue) and validates the re-weighting method.
 
The ratios between the differential curves obtained with each method are presented in the second inset.
This inset contains also the statistical uncertainty (yellow band) for the ratio of two independent SM samples. 
The compatibility of those two ratio plots with the expected statistical fluctuation validates our approach/code implementation. The first inset presents the ratio between the EFT and SM predictions. It shows that the method works correctly for quite small and quite large modifications of the differential distributions.
 
One can note that in the context of EFTs, the weight is linear in the dim-6 coupling\footnote{There would also be quadratic contribution if we include the squared matrix element associated to the dimension six operator.} therefore it is trivial to predict the weight from any value of the coupling as soon as the weights for two different values of the coupling are known. This property can be used to further speed up the computation of the weight.  
 
\subsection{$ZH$ associated production in the Effective Field Theory at NLO}

\begin{figure}[ht]
\begin{center}
\begin{minipage}{1.0\linewidth}
\centering
\includegraphics[width=1.0\linewidth]{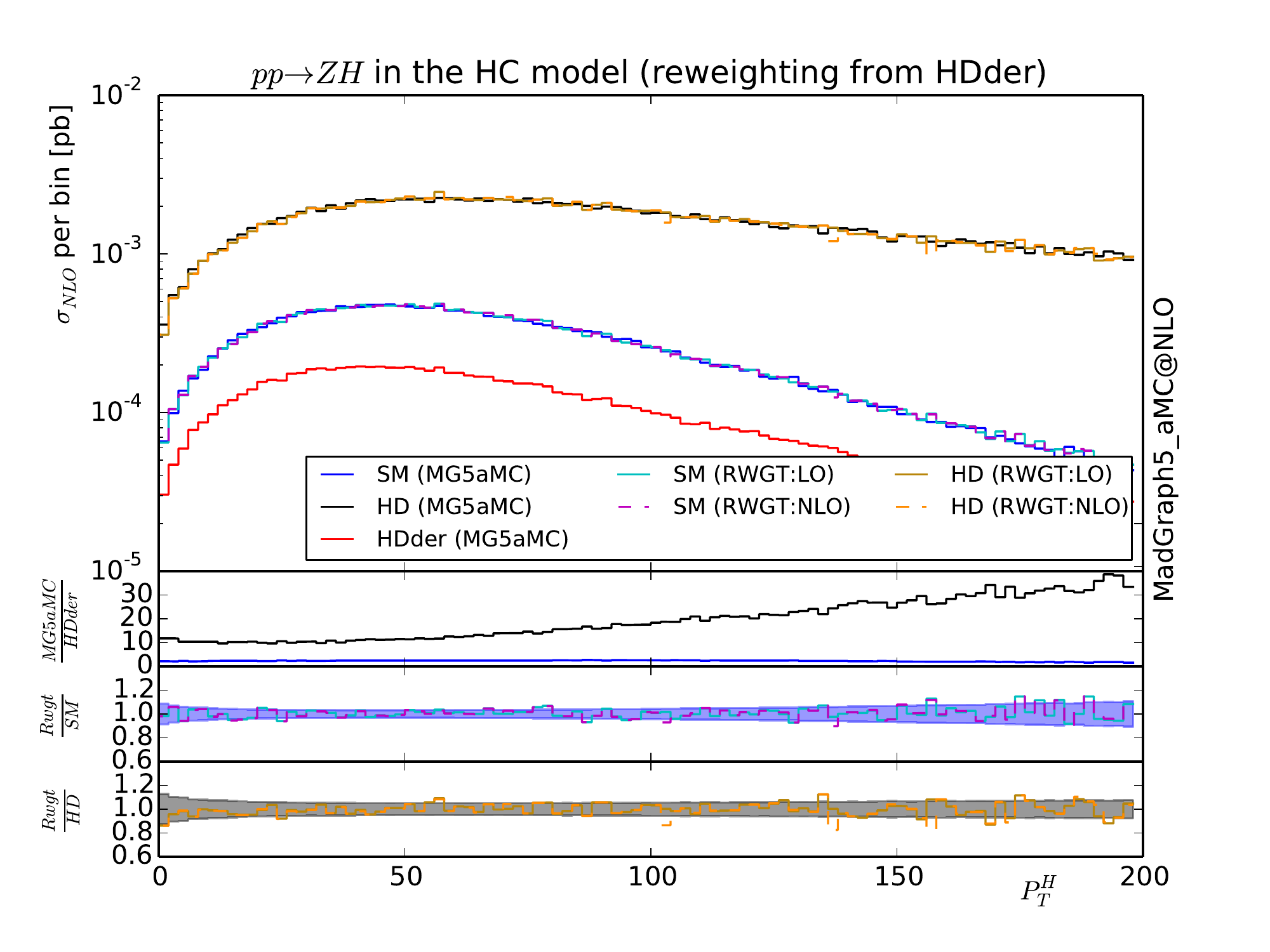}
\includegraphics[width=1.0\linewidth]{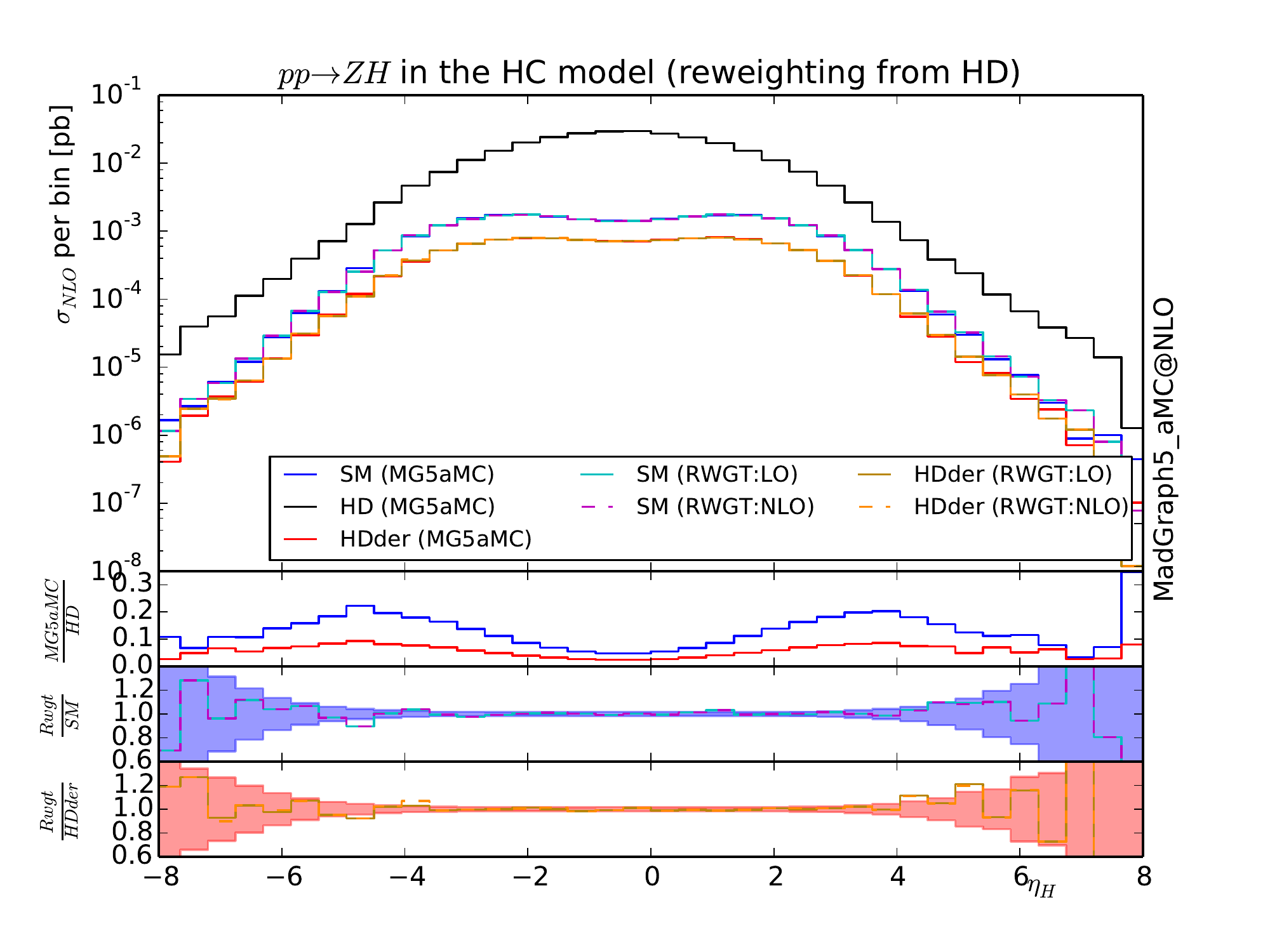}
\caption{\small{Differential cross-section for $p p \to Z H$ at 13 TeV LHC featuring both LO and NLO re-weighting methods.
Events have been showered with {\tt{Herwig6} }\cite{Corcella:2000bw}.
See text for details. 
}}
\label{figs:EFT_NLO}
\end{minipage}\hfill
\end{center}
\end{figure}

For our first NLO validation, we consider  the associated production of a $Z$ and $H$ boson in the  EFT as implemented in the Higgs Characterisation framework/model \cite{Demartin:2014fia}.
We use  two of the benchmarks introduced in \cite{Maltoni:2013sma}: HD and HDder. In more details, the effective Lagrangian relevant for this example is
\begin{eqnarray}
\mathcal{L}_{HD} &=& -\frac14 \frac1\Lambda \kappa_{HWW} Z_{\mu\nu}Z^{\mu\nu}H\\
 \mathcal{L}_{HDder} &=& -\frac1\Lambda \kappa_{H\partial Z} Z_\nu\partial_\mu Z^{\mu\nu}H + \nonumber\\
                                   &&(-\frac1\Lambda \kappa_{H\partial W}W^+_\nu\partial_\mu W^{-\mu\nu}H +h.c.)\,,
\end{eqnarray}
where $\Lambda$ is the high energy scale (set to $1 \textrm{TeV}$), $\kappa_{HWW}$, $\kappa_{H\partial Z}$, $\kappa_{H\partial W}$ are dimensionless couplings (set to one).
$H$ is the Higgs doublet field and $V_{\mu\nu} = \partial_\mu V_\nu -\partial_\nu V_\mu $; $V=Z, W^-,W^+$.

In Figure \ref{figs:EFT_NLO} we present the differential cross-section for the transverse momentum of the Higgs and for its rapidity. In both cases, we present the curve for the SM, HD and HDder benchmarks. For the transverse momentum, we start from an HDder sample of events and perform the re-weighting to the other scenarios.
While for the rapidity we present the plot where the original sample is the HD theory.
 Each re-weighted curve is then compared with a dedicated generation and the associated ratio plot is displayed below with the statistical uncertainty expected for the generation of two independent samples.
The agreement between the two is excellent for both the NLO accurate re-weighting and the Naive LO-like re-weighting.
In this case the NLO QCD effects factorise from the BSM ones and therefore the NLO accuracy of the Naive LO-like approach can only be spoiled by MC counter terms --which are as expected quite mild--.
One can also compare the statistical fluctuations between the \mgamc\ curves and the one obtained by re-weighting. If you look at the top plot (transverse momenta) for the HD case, it is clear that the statistical fluctuations are more pronounced for the curve obtained by re-weighting.
This is an example of enhancement of statistical uncertainty due to the re-weighting as discussed around Eq. \ref{eq:stat_uncertainty} since in the high $p_T$ region, the HDder is suppressed compare to the other theories under consideration (HD and SM).

\subsection{Effective Field Theory ($t\bar tZ$) at NLO}

In this second NLO example, we will use the EFT framework in the context of the top-quark \cite{Bylund:2016phk} and focus on the chromomagnetic operator:
\begin{equation}
\mathcal{O}_{tg}=y_t g_s(\bar{Q}\sigma^{\mu\nu}T^At)\tilde{\varphi}G_{\mu\nu}^A\,,
\end{equation}
where $Q$ is the third generation left-handed quark doublet,  $\varphi$ and $t$
are respectively the Higgs and top quark fields,  $g_s$ is the SM strong coupling constant,
$y_t$ is the top-Yukawa coupling  and $T^A$ is the $SU(3)$ generator.

\begin{figure}[ht]
\begin{center}
\begin{minipage}{1.0\linewidth}
\centering
\includegraphics[width=1.0\linewidth]{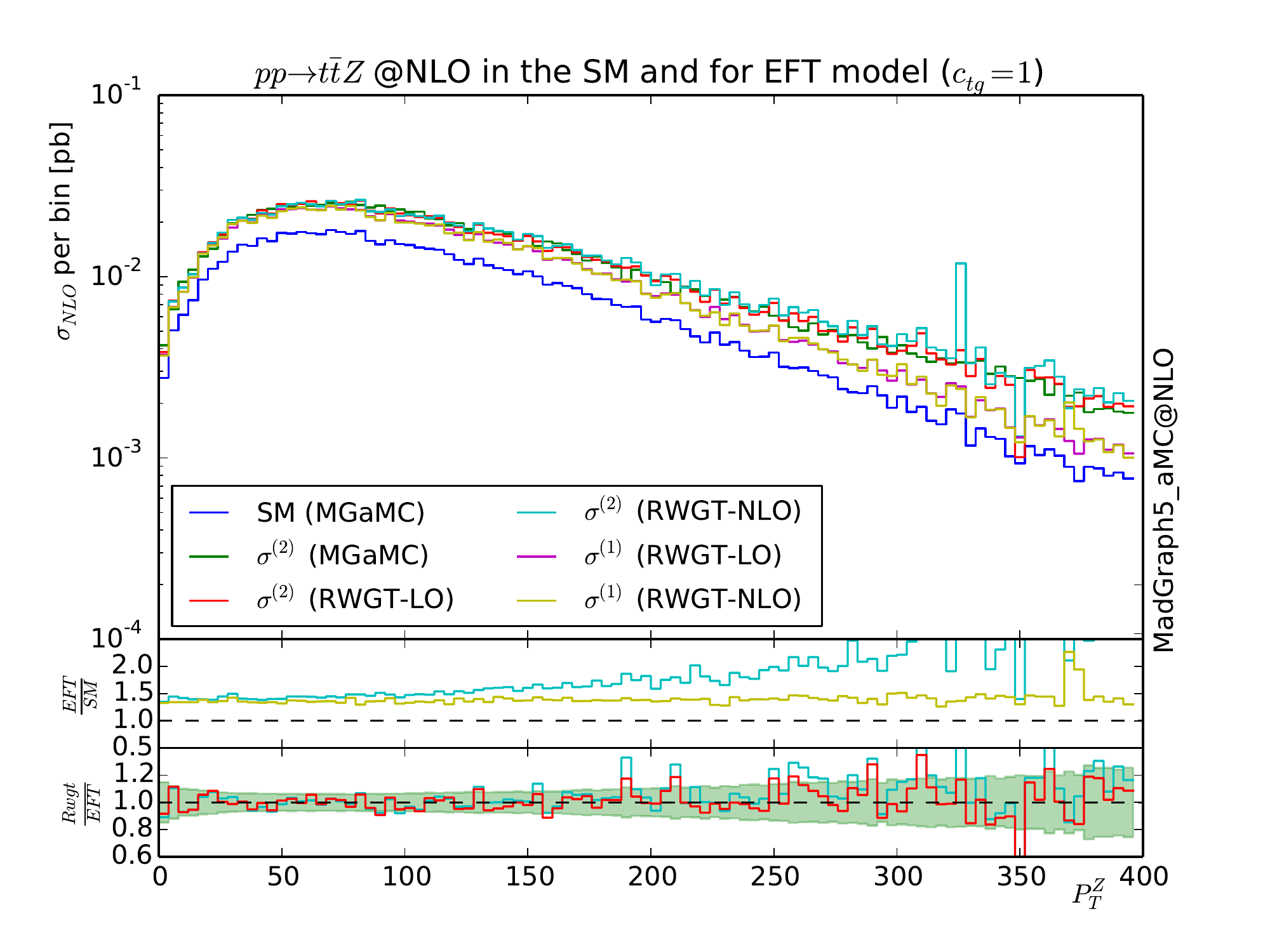}
\caption{\label{figs:TEFT}\small{
differential cross-section for $p p \to Z t \bar t$ at 13 TeV LHC featuring both LO and NLO re-weighting methods.
The shower have been performed with {\tt{Herwig6}} \cite{Corcella:2000bw}.
See text for details. 
}}
\label{figs:TEFT_NLO}
\end{minipage}\hfill
\end{center}
\end{figure}

In Figure \ref{figs:TEFT}, we present the transverse momentum of the $Z$ boson in the associated production of this boson with a top/anti-top quark pair.
We present the result for both the full matrix element squared (labelled $\sigma^{(2)}$) and for the SM contribution plus the interference with the dimension 6 operator only (labelled $\sigma^{(1)}$).
%Before this implementation, it was not possible to extract automatically such contribution from \mgamc.

As in the previous section, we present our prediction both via the Naive LO-like re-weighting method (RWGT-LO) and via the NLO accurate one (RWGT-NLO), The ratio to the SM curves are presented in the first inset while the ratio between our prediction and the direct computation in \mgamc\ for $\sigma^{(2)}$ is presented in the second inset. The green band represents the expected statistical uncertainty for the ratio of two \mgamc~samples. It is not possible to extract in automatic way the contribution of $\sigma^{(1)}$ from \mgamc\ and therefore we do not provide any comparison for this curve. As before, we observe  a case where the statistical uncertainties are enhanced by the re-weighting approach and where both the Naive LO-like and the NLO re-weighting provides similar results. In this case the theory do not factorise and the ratio of the virtual and of the Born are not expected to match. The nice agreement is explained by the small contribution of the virtual and, once again, by the mild effect of the MC counter terms.

\subsection{Higgs plus one jet production at LO and NLO order}

\begin{figure}[ht]
\begin{center}
\begin{minipage}{1.0\linewidth}
\centering
\includegraphics[width=1.0\linewidth]{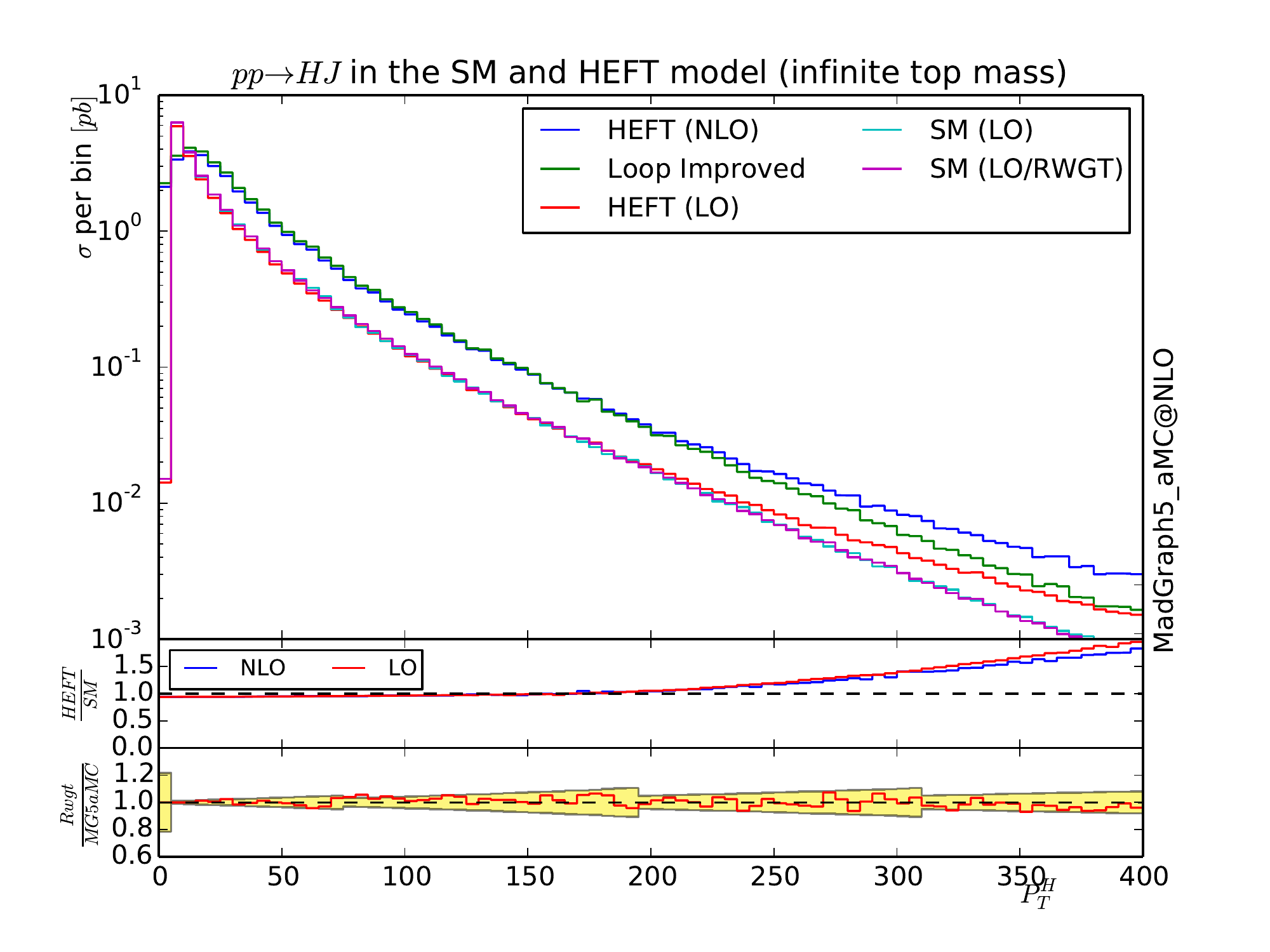}
\caption{\small{
Differential cross-section of the Higgs transverse momentum in the heavy top mass limit (both LO and NLO) re-weighted to include the finite top mass effect.
This is compare to the loop-induced processes (LO). 
The shower have been performed with {\tt{Herwig6}} \cite{Corcella:2000bw}. See text for details.
}}
\label{figs:HJ_NLO}
\end{minipage}\hfill
\end{center}
\end{figure}

In this last example, we will present results for the associated production of a SM Higgs with one jet.
In Figure \ref{figs:HJ_NLO}, we present the transverse momentum of the Higgs at both LO and NLO accuracy.  For the LO case, we present three curves. The first one is the curved obtained within the heft model \cite{Alwall:2007st} featuring the dimension five operator obtained by integrating out the top quark (HEFT LO). The second line (SM LO/RWGT) is the one obtained by re-weighting the previous curve by the full one loop matrix element squared which contain the complete top-quark mass dependence.
 The last LO curve is the one obtained via direct integration of the one-loop amplitude squared by \mgamc~\cite{Hirschi:2015iia} (SM LO).
At NLO accuracy, we have the curve in the infinite top mass limit (HEFT NLO) using the Higgs characterization model \cite{Maltoni:2013sma}.
This sample is then re-weighted by the full-loop (Loop Improved) following the loop-improved method presented in the previous section. It is so far not possible to compute the NLO contribution directly in order to compare the accuracy of such method.

The first inset presents the ratio at LO and NLO of the infinite top mass limit over the full theory. For the NLO case, the full theory is approximated by the loop-improved method. The two ratios are very similar showing that the loop-improved method re-introduces the top-mass effects in a sensible way. The second inset presents the ratio between the re-weighting and the direct approach in the LO case, the statistical uncertainty of the ratio of two independent SM sample is presented by the yellow band. His bumpy shape is due to the use of multiple samples with different cuts to decrease the statistical uncertainty. This ratio plot fully validates the re-weighting in the case of the LO curves.

\section{Conclusion}
\label{sec:conclusion}

We have presented the implementation of several methods that can be used for re-weighting  LO and NLO samples and discuss the associated intrinsic limitations.
We have released a new version of \mgamc~that allows the users to employe the various re-weighting methods presented in this paper  in a fully automatic and user-friendly way. In particular we have introduced for the first time an NLO accurate re-weighting method and compared it with the approximate methods available in the literature.
Other re-weighting methods like the Naive LO-like and the loop-improved are for the first time available in a public code.

The comparison between the various methods shows that the approximate method (the Naive LO-like re-weighting) performs a satisfactory job. This indicates that the non locality of the MC counter terms is often more a theoretical problem than a contribution spoiling the NLO accuracy of the Naive LO-like re-weighting. 
Therefore the Naive LO-like re-weighting should be a good approximation  in a quite large class of model/observable either when  the virtual contribution is sub-dominant and/or when the effect of the BSM physics factorises. 
Consequently, we recommend phenomenologist  to first test the Naive LO-like re-weighting and in case of loss of accuracy move forward to the slower NLO method.
On the other hand for mass production at the LHC, where the samples are often used for more than one study, we recommend to always use the NLO accurate method.
 
The framework introduced here is flexible enough to accommodate different types of re-weighting approaches. In particular we plan to extend this work in the direction of systematics computation
allowing to modify not only the matrix element but also the scale scheme, the pdf set, and so on,  both at LO and NLO. Compared to \cite{Frederix:2011ss} it will allow to perform such  re-weighting independently of the event generation which will be extremely useful to evaluate the effect of a new PDF set/test a new scale scheme on existing samples.

\appendix

\section{Theoretical proof}
\label{annexe:proof}
\subsection{Unweighting}

In order to have a formal proof that the unweighting procedure can commute with the re-weighting, we first have to formalize the procedure.
Following the convention adopted in the previous sections, a standard Monte-Carlo integration is:
\begin{eqnarray}
\sigma_{orig}           &=& \sum_{i=1}^N f_1(x_1^i,\mu_F)\cdot f_2(x_2^i,\mu_F)\cdot|M_{orig}^i|^2\cdot d\Omega^i \label{eq:wgt_def2}\\
                                &\equiv& \sum_{i=1}^N W^i_{orig},
\end{eqnarray}

To get an unweighted sample, we first need to multiply and divide this expression by $\max_i (W^i_{orig}) $:
\begin{eqnarray}
\sigma_{orig} &=& \max_i (W^i_{orig})  \sum_{i=1}^N \frac{W^i_{orig}}{\max_i (W^i_{orig})},
\end{eqnarray}
Finally, the term $\frac{W^i_{orig}}{\max_i (W^i_{orig})}$ can be re-interpretted as a probability to accept/reject the phase-space point.\footnote{For non definite positive quantity the same idea holds by using $\max_i (|W^i_{orig}|)$.}

 By randomly selecting a sub-sample of phase-space points with that probability, we reduce significantly the sample size.
 Additionally, all the remaining  events have the same weight ($\max_i (W^i_{orig})$) and the associated distribution of events follows the physical distributions.
 
\begin{equation}
\sigma_{orig} \approx  \max_i (W^i_{orig})  \sum_{i=1}^N Acc_i = \max_i (W^i_{orig}) * N_{acc}.
\end{equation}
where $Acc_i$ is either $0$ or $1$  depending on whether the event was kept or rejected following the $\frac{W^i_{orig}}{ \max_i(W^i_{orig})}$ probability distribution.

Let's now proof that the re-weighting works on a unweighted sample, by doing the same for a second theory. But instead of multiplying and dividing by $\max_i(W^i_{new})$ we will use the maximum weight of the original theory:

\begin{eqnarray}
\sigma_{new} &=& \sum_{i=1}^N W^i_{new}, \\
                         &=& \max_i(W^i_{orig})  \sum_{i=1}^N \frac{W^i_{new}}{ \max_i(W^i_{orig})}.
\end{eqnarray}
Since $W^i_{new} = \frac{|M^i_{new}|^2}{|M^i_{old}|^2}W^i_{old}$ (See Eq. \ref{eq:wgt_def2}), this is equal to
\begin{equation}
\sigma_{new}                = \max_i(W^i_{orig})  \sum_{i=1}^N  \frac{W^i_{new}}{W^i_{orig}}\frac{W^i_{orig}}{ \max_i(W^i_{orig})}.
\end{equation}
We recover in that equation the same ratio which was used to unweight the original theory. We can therefore select the same sub-sample of events and just re-weight them by the ratio of the matrix element squared.

%%%%%%%%%%%%%%%%%%%%%%%%%%%%%%%%%%%%%%%%%%%%%%%%%%%%%%%%%%%%%%%%%%%
\subsection{statistical uncertainty from an un-weighted sample}

One can notice that the estimated uncertainty can not be obtained via re-weighting for an unweighted sample due to the non linear dependence in the matrix element squared.

We will show in this section what needs to be done in order to build an estimator of the variance from a re-weighted sample.  Following the idea of the unweighting procedure, we can rewrite the standard estimator of the variance by:
\begin{eqnarray}
\frac{\Delta \sigma_{orig}^2}{N} &=& \sum_{i=1}^N {W^i_{orig}}^2 - \frac1N \left(\sum_{i=1}^N W^i_{orig}\right)^2 \\
                                    &=&  \max_i(W^i_{orig})^2 \left[\sum_{i=1}^N \frac{W^i_{orig}}{ \max_i(W^i_{orig})}\frac{W^i_{orig}}{ \max_i(W^i_{orig})}\right. \nonumber\\
                                    &\phantom{ee}-&  \left.\frac1N \left(\sum_{i=1}^N \frac{W^i_{orig}}{ \max_i(W^i_{orig})}\right)^2  \right],
\end{eqnarray}
As for the unweighting case, we can re-interpret the  ratio $\frac{W^i_{orig}}{ \max_i(W^i_{orig})}\equiv P^{orig}_{acc,i}$ as the probability to keep the event during the unweighting procedure.
Therefore after the event unweighting the equation can be read as:
\begin{eqnarray}                                    
 \frac{\Delta \sigma_{orig}^2}{N} &\approx& \max_i(W^i_{orig})^2  \left[\sum_{i=1}^{N_{acc}}  \frac{W^i_{orig}}{ \max_i(W^i_{orig})} -  \frac{N_{acc}^2}N\right]
\end{eqnarray}
In this case,  a dependence remains in the unweighting probability as well as in the number of generated and accepted events.
If those informations were kept during the unweighting procedure it would be possible to construct the above estimator of the variance. The re-weighting of such information is then possible and 
one can construct such an  estimator for any re-weighted sample:
\begin{eqnarray}
\frac{\Delta \sigma_{new}^2}{N} &\approx&  \max_i(W^i_{orig})^2 \left[\sum_{i=1}^{N_{acc}} P^{orig}_{acc,i} \left(\frac{|M_{new}|^2}{|M_{orig}|^2}\right)^2\right.\nonumber\\
                                                   && \phantom{  \max_i(W^i_{orig})^2 }-\left. \frac1N \left(\sum_{i=1}^{N_{acc} } \frac{|M_{new}|^2}{|M_{orig}|^2} \right)^2\right].
\end{eqnarray}
Note that in presence of multi-channel integration such information need to be provided for each channel individually.

This method is currently not implemented in \mgamc\ but we plan to include it in a near future and study the accuracy of such an estimator.

%%%%%%%%%%%%%%%%%%%%%%%%%%%%%%%%%%%%%%%%%%%%%%%%%%%%%%%%%%%%%%%%%%%
\subsection{NLO-reweighting}

In order to proof that the re-weighting proposed in Eq. \ref{eq:NLO_VIRT} is correct we first need to formalise the loop integration method.
We will use in this section a simplified notation such that
\begin{eqnarray}
\sigma^{soft}_{orig} & = &  \sum_{i=1}^N (B^i_{orig} + V^i_{orig} + C^i_{orig}) \\
                                & \equiv & \sigma^{soft,B}_{orig}  +  \sum_{i=1}^N C^i_{orig}.
\end{eqnarray}
Where $B$, $V$, $C$ represents respectively the Born, the virtual and the counter terms contribution. 
Since the counter terms do not play any role in this optimisation procedure (and have a natural re-weighting) we will focus on the $ \sigma^{soft,B}_{orig} $ pieces:
In this simplified formalism the phase-space optimisation method can be written has (see Eq. \ref{eq:virttrick}):
\begin{eqnarray}
\sigma^{soft,B}_{orig}      & = & \sum_{i=1}^N (B^i_{orig} + V^i_{orig})\\
                                    & = & \sum_{i=1}^N (B^i_{orig} + \kappa_{orig}*B^i_{orig}))\nonumber \\ 
                                    &  & + \sum_{i=1}^N (V^i_{orig} - \kappa_{orig}*B^i_{orig} )\label{eq:app:proof:vt_1}\\
                                    & \simeq & \sum_{i=1}^N (B^i_{orig} + \kappa_{orig}*B^i_{orig})) \nonumber  \\
                                    &&+ \sum_{j=1}^{N/k} k* (V^j_{orig} - \kappa_{orig}*B^j_{orig} ).    \label{eq:vt}                            
\end{eqnarray}
In those equations, we first (Eq.~\ref{eq:app:proof:vt_1}) add and subtract the approximant of the virtual: $ \kappa_{orig}*B^i_{orig}$, while in the second equation we integrate on different statistics the two pieces of the sum. We run $k$ times less phase-space point in the second and therefore have to multiply it by  the factor $k$.

If we want to use the re-weighting on the sample generated via this method, we have to apply the same method with the same value of $\kappa_{orig}$

\begin{eqnarray}
\sigma^{soft,B}_{new}      & = & \sum_{i=1}^N (B^i_{new} + \kappa_{orig}*B^i_{new})\nonumber \\ 
                                     &  & + k*\sum_{i=1}^{N/k}  (V^i_{new} - \kappa_{orig}*B^i_{new} ).                                 
\end{eqnarray}

Inspired by Eq. \ref{eq:NLO_VIRT}, we will multiply all those terms by the identity factor $1 = \frac{B^i_{orig} + V^i_{orig}}{B^i_{orig}+V^i_{orig}}$:                                                                                                                                               
\begin{eqnarray}
\sigma^{soft,B}_{new}  & = & \sum_{i=1}^N (B^i_{new} + \kappa_{orig}*B^i_{new}) \frac{B^i_{orig} + V^i_{orig}}{B^i_{orig}+V^i_{orig}}\nonumber \\ 
                                     &  & + k*\sum_{i=1}^{N/k} (V^i_{new} - \kappa_{orig}*B^i_{new}) \frac{B^i_{orig} + V^i_{orig}}{B^i_{orig}+V^i_{orig}}.
\end{eqnarray}  

We can rewrite the expression as the expected re-weighting formula plus some rest-over
\begin{eqnarray}  
\sigma^{soft,B}_{new}   & = & \sum_{i=1}^N   \frac{B^i_{new} + V^i_{new}}{B^i_{orig}+V^i_{orig}} (B^i_{orig}+\kappa_{orig} B^i_{orig}) \nonumber \\
                                     & &   - \sum_{i=1}^N  \frac{(1+\kappa_{orig})}{B^i_{orig}+V^i_{orig}}  ( V^i_{orig}B^i_{new}- V^i_{new}B^i_{orig})   \nonumber \\
                                     & &  - k* \sum_{i=1}^{N/k}   \frac{(1+\kappa_{orig}) }{B^i_{orig}+V^i_{orig}} (V^i_{new}B^i_{orig} - V^i_{orig}B^i_{new}) \nonumber \\
                                     & &  +k*\sum_{i=1}^{N/k}  \frac{B^i_{new} + V^i_{new}}{B^i_{orig}+V^i_{orig}} (V^i_{orig}-\kappa_{orig} B^i_{orig}). \label{eq:expanded}
\end{eqnarray}
If the same phase-space sampling is used for both parts ($k=1$) then the second and third lines cancel. The remaining lines correspond to the re-weighting of Eq. \ref{eq:NLO_VIRT}.
If both integral are sampled in a different way ($k\neq 1$), then the cancellation is not exact but should still occur for large  enough samples. Therefore this optimization method introduces a new contribution to the statistical uncertainty.

\begin{acknowledgements}

I would like to thank all the authors of \mgamc\ for their discussions, help and support at many stages of this project.
I would like also to thank C. Degrande, R. Frederix and F. Maltoni to have read and comment on this manuscript,
E. Vryonidou, F. DeMartin, I.Tsinikos, V. Hirschi for their help during the validation of this implementation.
 O.M. is supported by a Durham International Junior Research Fellowship. 
This work is supported in part by the IISN "MadGraph" convention 4.4511.10,
by the Belgian Federal Science Policy Office through the Interuniversity Attraction Pole
P7/37, by the European Union as part of the FP7 Marie Curie Initial Training Network
MCnetITN (PITN-GA-2012-315877), and by the ERC grant 291377 ÔLHCtheory: Theoretical 
predictions and analyses of LHC physics: advancing the precision frontierÕ.

\end{acknowledgements}

% BibTeX users please use one of
%\bibliographystyle{spbasic}      % basic style, author-year citations
%\bibliographystyle{spmpsci}      % mathematics and physical sciences
\bibliographystyle{spphys}       % APS-like style for physics
\bibliography{rwgt_biblio}   % name your BibTeX data base

% Non-BibTeX users please use
%\begin{thebibliography}{}
%
% and use \bibitem to create references. Consult the Instructions
% for authors for reference list style.
%
%\bibitem{RefJ}
% Format for Journal Reference
%Author, Article title, Journal, Volume, page numbers (year)
% Format for books
%\bibitem{RefB}
%Author, Book title, page numbers. Publisher, place (year)
% etc
%\end{thebibliography}

\end{document}